\documentclass[aps,twocolumn,pra,tightenlines,floatfix,showpacs]{revtex4}

\usepackage[dvips]{graphicx}
\usepackage[english]{babel}
\usepackage{amsmath}
\usepackage{amssymb}
\usepackage{times}

\newcommand{\bs}{\begin{split}}
\newcommand{\es}{\end{split}}
\newcommand{\be}{\begin{equation}}
\newcommand{\ee}{\end{equation}}
\newcommand{\ba}{\begin{eqnarray}}
\newcommand{\ea}{\end{eqnarray}}

\newcommand{\Ek}{E_{\mathbf{k}}}

\begin{document}

\title{Superfluid phase diagrams of trapped Fermi gases with population
  imbalance}

\author{Chih-Chun Chien, Qijin Chen, Yan He, and K. Levin}

\affiliation{James Franck Institute and Department of Physics,
 University of Chicago, Chicago, Illinois 60637}

\date{\today}

\begin{abstract}
  We present phase diagrams for population imbalanced, trapped Fermi
  superfluids near unitarity. In addition to providing quantitative
  values for the superfluid transition temperature, the pairing onset
  temperature and the transition line (separating the Sarma and phase
  separation regimes), we study experimental signatures of these
  transitions based on density profiles and density differences at the
  center. Predictions on the BCS side of resonance show unexpected
  behavior, which should be searched for experimentally.
\end{abstract}

\pacs{03.75.Hh, 03.75.Ss, 74.20.-z \hfill \textsf{\textbf{cond-mat/0612103}}}

\maketitle


The study of ultracold trapped Fermi gases, as they vary from BCS to
Bose-Einstein condensation (BEC) is a rapidly exploding subject which is
defining new directions in condensed matter and atomic physics.  With
the recent focus on population imbalanced superfluids
\cite{ZSSK06,Rice1,ZSSK206,Rice2}, this body of work has also captured
the attention of physicists in other disciplines \cite{FGLW05}.  Indeed,
it is hard in recent years to find a subfield of physics which has such
a broad appeal across the different sub-specialties.  The goal of the
present paper is to address the multiple superfluid and normal phases
which appear in these trapped gases and which are viewed as possible
prototypes for high temperature superconductors \cite{Reviews}, as well
as quark and nuclear matter.  The various phases we contemplate become
stable or unstable as one alters the populations of the two different
spin species or changes the temperature $T$.  There has been a very
extensive theoretical literature on this subject which is almost
exclusively confined to $T=0$ \cite{SR06,Mueller,HS06}.  Our emphasis
here, is on finite temperature effects \cite{Reviews,0608662}, which are
essential in order to address the actual experimental situation.  We do
this here along with presenting predictions for new experiments.

The fermionic superfluid ground state wavefunction in the presence of
BCS-BEC crossover effects (with population imbalance) is almost
universally assumed \cite{SR06,Mueller,HS06} to be of the BCS-Leggett
form.  The excitations of this ground state are expected to be
particularly interesting in the intermediate or unitary regime where
they consist of non-condensed pairs as well as gapped fermions.  While
these finite momentum pair excitations are frequently omitted in the
literature \cite{Machida2,Stoof06b}, we stress that they play an
essential role.  Only with their inclusion will one find quantitatively
meaningful estimates of $T_c$; this temperature is reduced dramatically
(relative to a naive mean field estimate) as a result of the opening of
an excitation or pairing gap well above the transition.  Indeed, the
separation of the pairing onset temperature $T^*$ and the phase
coherence temperature $T_c$ is well documented in experiments which are
consistent with a ``pseudogap'' (or normal state excitation gap) in the
unitary trapped gases \cite{Reviews}.

Without the contribution of non-condensed pairs \cite{JS5,Reviews} one
often finds bimodal particle distributions at $p=0$.  Therefore,
theories which ignore these pairs \cite{Machida2,Stoof06b} may not be
applicable for establishing bimodal distributions associated with
population imbalance.  Pairing fluctuations were partially included
(through the number equation) in Ref.  \cite{Parish06} based on the
Nozie\`res--Schmitt-Rink (NSR) scheme \cite{NSR} but in the absence of a
trap. However, the finite $T$ NSR treatment is not designed to be
consistent with the standard ground state \cite{SR06,SM06,HS06}.
In the presence of population imbalance, it is now well known that one
has to consider homogeneous phases (which we refer to as the ``Sarma'') and
Larkin-Ovchinnokov-Fulde-Ferrell (LOFF) phases \cite{FFLO} as well as
phase separation \cite{Caldas04}.  In addition to these competing
phases, the normal phase may appear as a highly correlated or paired
state without phase coherence, or as a simple Fermi gas (unpaired)
phase.

We begin with the (local) thermodynamical potential (per unit volume)
for the allowed states (excluding the more complicated LOFF phase which
appears to be of less interest near unitarity and at all but the lowest
temperatures \cite{LOFF1}). We focus on a two-component Fermi gas in a
harmonic trap potential.  For a normal or superfluid phase in which
pairing correlations are present the thermodynamical potential
($\Omega_{tot}$) consists of two highly inter-dependent contributions:
from the gapped fermions ($\Omega_f$) and the non-condensed pairs or
bosons ($\Omega_b$)
\begin{eqnarray}
  & &\Omega_{tot}=\Omega_f+\Omega_b\\
  & &\Omega_{f}=-\frac{\Delta^2}{U}+\sum_{\mathbf{k}}(\xi_{k}-E_{k})
  -T\sum_{\mathbf{k},\sigma}\ln{\left(1+e^{-E_{k\sigma}/T}\right)} \nonumber\\
  & &\Omega_{b}=
  Z\mu_{pair}\Delta^{2}_{pg}+T\sum_{\mathbf{q}}
  \ln{\left(1-e^{-\tilde{\Omega}_{q}/T}\right)}.\nonumber  
\end{eqnarray}
Competing with this phase is the free Fermi gas phase which has
thermodynamical potential 
$\Omega_{free}=-T\sum_{\mathbf{k},
\sigma}\ln{\left(1+e^{-\xi_{k\sigma}/T}\right)}$.
The (gapped) fermion and pair dispersions are given respectively by
$\Ek=\sqrt{\xi_{\bf k}^{2}+\Delta^{2}}$ for a contact (short-range)
pairing interaction with strength $U$ and $\tilde{\Omega}_q = q^2/2M^* -
\mu_{pair}$, where $M^*$ and $\mu_{pair}$ are the effective mass and
chemical potential of the pairs. Both $M^*$ and $Z$ are obtainable from
a microscopic $T$-matrix approach\cite{0608662}.  Here
$E_{\mathbf{k}\sigma} = \Ek \mp h$ and $\xi_{\mathbf{k}\sigma} =
\xi_{\mathbf{k}} \mp h$ for spin $\sigma = \uparrow,\downarrow$,
respectively, where $\xi_{\mathbf{k}} = k^2/2m -\mu$, $k_B=\hbar =1$
with $\mu = (\mu_\uparrow + \mu_\downarrow)/2$ and $h = (\mu_\uparrow -
\mu_\downarrow)/2$.
%

We distinguish the order parameter $\Delta_{sc}$ from the total gap
$\Delta$.  Non-condensed pairs are associated with a pseudogap
contribution, $\Delta_{pg}$, to the total gap via
$\Delta^{2}=\Delta_{sc}^{2}+\Delta_{pg}^{2}$.  The form of $\Omega_{b}$
contains a free boson-like contribution; the pair condensate does not
contribute to $\Omega_{b}$ directly.  Although it is not a necessary
assumption, in order to simplify the formal description we assume that
$M^*$ depends on $T$ only, as is reasonably consistent with our
microscopic theory \cite{0608662} and $\mu_{pair}$ is a function of
$\Delta$ and $T$ only.  In the summation over boson momentum
$\mathbf{q}$, we will impose a cutoff $q<q_{c}$, where $q_c$ is the
minimum value of $q$ which satisfies
$E_{k}+\xi_{\mathbf{k+q}}-\tilde{\Omega}_{q}=0$ for at least one value
of $\mathbf{k}$;
above this momentum, pairs become diffusive by decay into the
particle-particle continuum. It should be noted that $\Omega_f$ and
$\Omega_b$ couple only via $\mu_{pair}$.  Above $T_c$ this leads to an
extra contribution to the gap equation.

Our self-consistent equations can be expressed as variational conditions
on $\Omega_{tot}$, which yield equations for the total excitation gap
$\Delta$, the pseudogap $\Delta_{pg}$, the fermion number density $n$
and population imbalance $\delta n$, respectively:
$$\frac{\partial\Omega_{tot}}{\partial\Delta}=0;\quad\!
\frac{\partial\Omega_{tot}}{\partial\mu_{pair}}=0; \quad\!
-\frac{\partial\Omega_{tot}}{\partial \mu}= n; \quad\!
-\frac{\partial\Omega_{tot}}{\partial h}= \delta n.
$$
Importantly, below $T_c$ we have $\mu_{pair}=0$, and above $T_c$ we have
$\Delta= \Delta_{pg}$.  These equations reduce to the usual temperature
dependent BCS-like equations for the gap and number equations in the
literature.  The effect of the pseudogap, which is new to the present
formalism, has also been extensively discussed in earlier papers
\cite{Reviews,0608662}, although here we have presented it in a
different, but equivalent fashion.

In a trap, $T$ and $h$ are independent of the radial position.  The
distribution of the local $\mu(\mathbf{r})$ is determined by the force
balance equation, $-\nabla \bar{p} =n\nabla V_{ext}$, where $\bar{p} =
-\Omega_{tot}$ is the pressure and
$V_{ext}=\frac{1}{2}m\bar{\omega}^2r^2$ is the harmonic trap potential
with mean angular frequency $\bar{\omega}$.  Using the number equation
$n=-\partial\Omega_{tot}/\partial\mu$, we obtain $\nabla
\mu(\mathbf{r})=-\nabla V_{ext}(\mathbf{r})$, or
$\mu(\mathbf{r})=\mu_0-V_{ext}(\mathbf{r})$, where $\mu_0 \equiv \mu(0)$
and $V_{ext}(0)=0$. This shows that the force balance condition
naturally leads to the usual local density approximation (LDA).
To ensure meaningful comparisons between states under different
conditions, we fix the total particle number $N=\int d^3 r\, n(r)$ and
the number difference $\delta N$.  The Fermi energy $E_{F}=T_{F}\equiv
k_F^{2}/2m$ is defined as that of a non-interacting system with the same
$N = N_\uparrow + N_\downarrow$ at the polarization $p\equiv \delta N/N
=0$.  We assume $N_\uparrow > N_\downarrow$ so that $h>0$.

The physical state corresponds to the minimum of $\Omega_{tot}$ and
$\Omega_{free}$.  At a particular trap radius when (and if) these become
equal, the system will have a first order transition from a state in
which there is pairing (but not necessarily superfluidity) to an
unpaired Fermi gas phase.  We assume an infinitely thin interface (i.e.,
domain wall) for this phase separation.  There is as yet no complete
microscopic theory for the interface energy, so that we do not include
it here.
The philosophy behind our treatment of phase separation is very similar
to that in previous $T=0$ papers \cite{Mueller,Chevy} except that we
include the important effects of temperature (in a fashion consistent
with the well studied BCS-Leggett ground state).  Across the interface,
we require thermal, chemical, and mechanical equilibrium so that $T$,
$\mu_\sigma$ and $\bar{p}$ are continuous.

\begin{figure}
\centerline{\includegraphics[clip,width=3.2in] {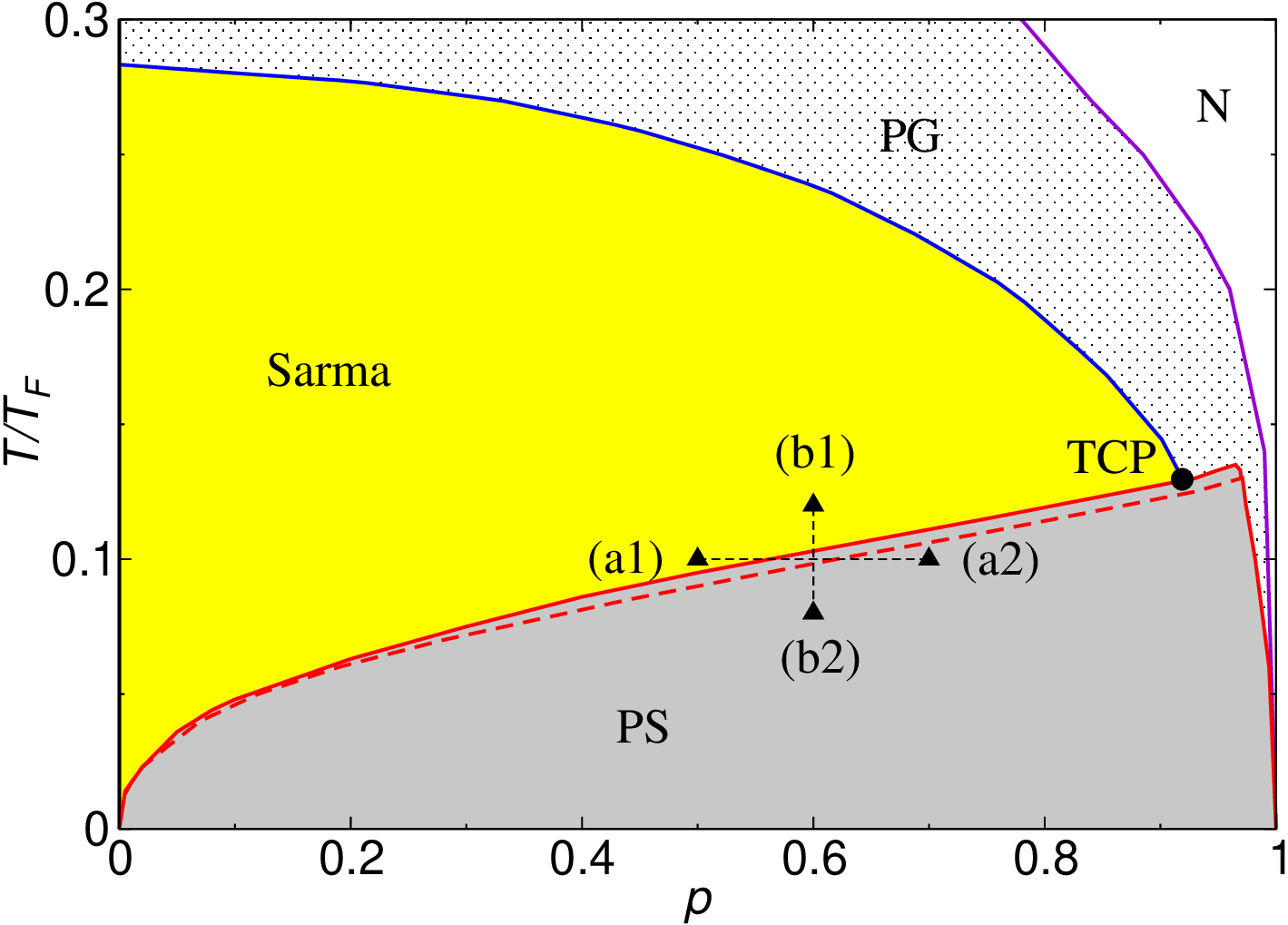}}
\caption{(Color online) Phase diagram of a population-imbalanced Fermi
  gas in a harmonic trap at unitarity. The solid lines separate
  different phases. Above the (red) dashed line but within the PS phase,
  the superfluid core does not touch the domain wall.  Here ``PG''
  indicates the pseudogapped normal phase.  The black dot labeled
  ``TCP'' indicates the tricritical point.  The four points indicated by
  the triangles labeled (a1)-(b2) correspond to the density profiles in
  Fig.~\ref{fig:profile}. }
\label{fig:TP_trap}
\end{figure}

Fig.~\ref{fig:TP_trap} shows the phase diagram at unitarity for a
population-imbalanced Fermi gas in a trap.  Here $1/k_{F}a=0$, where $a$
is the $s$-wave scattering length between fermions.
Phase separation (labeled PS) occupies the lower $T$ portion of
the phase diagram, where the gap $\Delta$ jumps abruptly to zero at some
trap radius.  At intermediate $T$, there is a (yellow-shaded) Sarma
phase, where $\Delta$ vanishes continuously within the trap.  It evolves
into a (dotted) pseudogap (PG) phase as the superfluid core vanishes at
higher $T$.  A normal (N) phase without pairing  always
exists at even higher $T$.
In contrast with earlier work \cite{Machida2,Stoof06b}, here we include
pair fluctuations or non-condensed pairs, which are essential in order
to obtain quantitatively correct values for $T_c$ as well as for the
finite temperature particle density profiles \cite{JS5,Reviews}.

We stress that with or without the trap, the PS phase is the ground
state at unitarity for any polarization.  The differences between the
trap and homogeneous phase diagrams are principally quantitative.  The
finite temperature boundary between the Sarma and PS phase is shifted as
a result of the trap potential, allowing the Sarma state to exist at low
$T$ and low polarizations.
The location of the tricritical point (TCP) also shifts from low
polarization in the uniform case (not shown) to high polarization in the
trapped case.  Away from $p \equiv 0$ there is a minimum temperature
required to arrive at this Sarma phase, as a consequence of the well
documented negative $T=0$ superfluid density.

\begin{figure}
\centerline{\includegraphics[clip,width=3.2in]{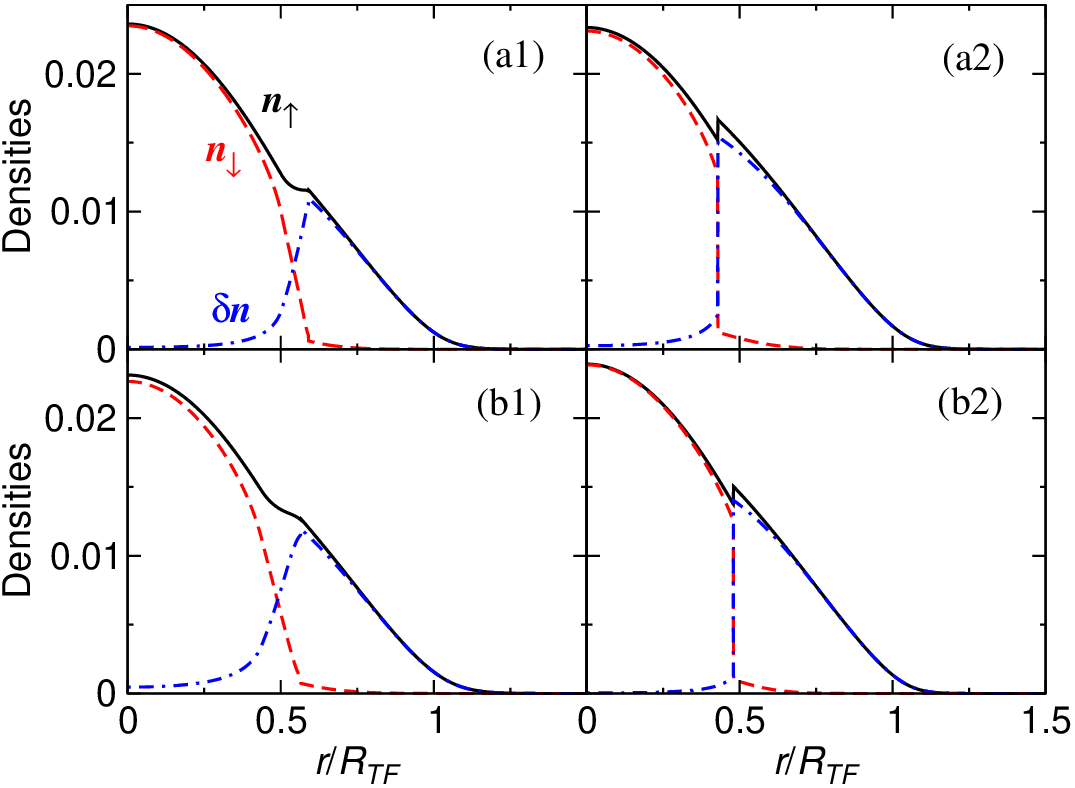}}
\caption{(Color online) Three-dimensional density profiles corresponding
  to the four points labeled in the phase diagram at unitarity shown in
  Fig.~\ref{fig:TP_trap}: (a1) $T=0.1T_{F}$, $p=0.5$. (a2) $T=0.1T_{F}$,
  $p=0.7$.  (b1) $T=0.12T_{F}$, $p=0.6$. (b2) $T=0.08T_{F}$, $p=0.6$.
  The (black) solid, (red) dashed, and (blue) dot-dashed lines
  correspond to $n_{\uparrow}$, $n_{\downarrow}$, and $\delta n$,
  respectively.  Here $R_{TF}=\sqrt{2E_{F}/m\bar{\omega}^{2}}$ is the
  Thomas-Fermi radius, and the units for density are $k_F^{-3}$.}
\label{fig:profile}
\end{figure}

In Fig.~\ref{fig:profile} we present four representative density
profiles which show the behavior on different sides of the transition
line between the PS and Sarma phases.  The (a1)-(a2) pair have the same
$T$ but different $p$, while the (b1)-(b2) pair have the same $p$ but
different $T$. We have chosen these points some distance from the
transition line in order to illustrate the differences. It should be
noted, however, as points (a1) and (b1) (which are within the Sarma
phase), move closer to the transition line, $\Delta$ drops rapidly (but
not discontinuously), as a precursor to abrupt phase separation.  The
transition from the pure Sarma phase to the PS phase is, thus, a
relatively smooth one.  These LDA-based calculations should apply to
situations where the trap geometry is reasonably isotropic, as for
example in the MIT experiments \cite{ZSSK06,ZSSK206} and where one might
be able to ignore surface energy contributions.  On the other hand, this
behavior appears at odds with recent experiments from Rice
\cite{Rice1,Rice2} which report pronounced changes in the aspect ratio
of the profile as the transition line is crossed.  These changes have
been attributed to extreme trap anisotropy and associated interface
energy effects \cite{Mueller06}.

\begin{figure}
\centerline{\includegraphics[clip,width=3.in]{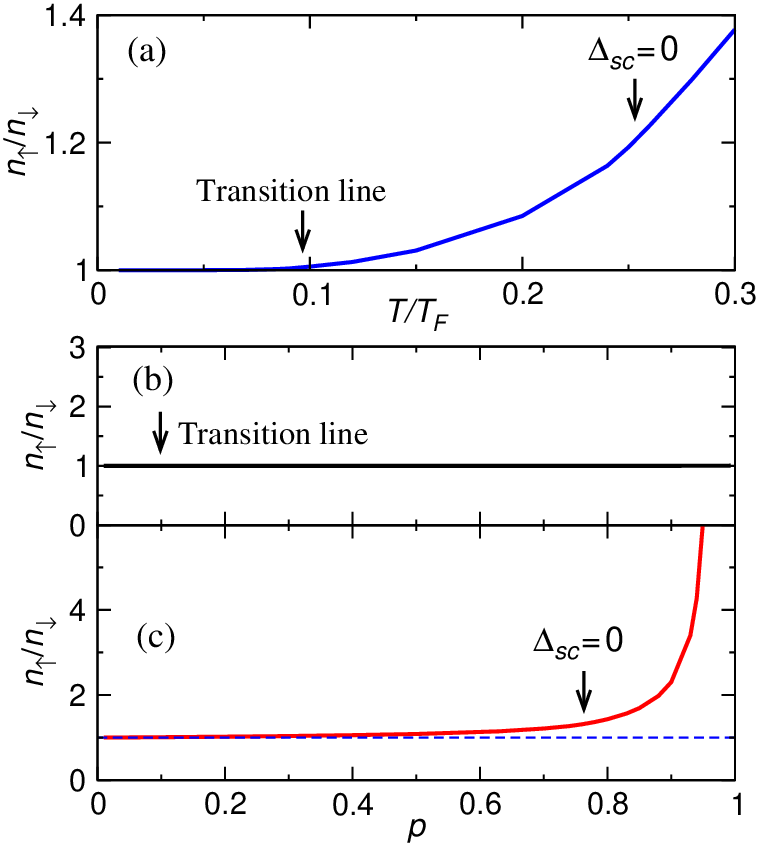}}
\caption{(Color online) Behavior of $n_{\uparrow}/n_{\downarrow}$ at the
  trap center at unitarity, as a function (a) of $T/T_{F}$ at fixed
  $p=0.5$, of $p$ at fixed (b) $T=0.05T_{F}$ and (c) $T=0.2T_{F}$. The
  (blue) dashed line in (c) indicates $n_{\uparrow}/n_{\downarrow}=1$.
  The arrows labeled ``Transition line'' indicate where the PS-Sarma
  transition occurs, and the arrows labeled ``$\Delta_{sc}=0$'' indicate
  where superfluid condensate disappears in the Sarma state.}
\label{fig:n1n2}
\end{figure}

In a different class of experiments it was proposed that by measuring
densities at the center of the trap as one sweeps $T$ or $p$, one can
infer when the Sarma-PS transition line is crossed \cite{MITPRL06} as
well as where superfluid transition [i.e., $T_c(p)$] occurs
\cite{Rice2}.  As in these experiments, we plot the ratio
$n_{\uparrow}/n_{\downarrow}$ at the trap center as a function of $T$
and $p$ in Fig.~\ref{fig:n1n2}.  As shown in Fig.~\ref{fig:n1n2}a, when
$p$ is fixed at a relatively high value while sweeping $T$, the curve
starts at $1$ in the PS regime at low $T$, and begins to increase when
the the transition line to the Sarma phase is crossed.
This behavior reflects the fact that Sarma states can accommodate higher
core polarizations than their PS counterparts.  In Fig.~\ref{fig:n1n2}b,
when instead $T$ is fixed below TCP while sweeping $p$, the curve is a
straight line independent of $p$, as observed experimentally
\cite{Rice2}.  Here one can infer, that the core has equal population in
both the Sarma state at low $T$ and the PS phase at any $(T,p)$.  This
figure also underlines the fact that PS states exist up to very high
polarizations ($p>0.99$).  Finally, when $T/T_{F}$ is higher than that
of the TCP (Fig.~\ref{fig:n1n2}c), the ratio increases rapidly after the
superfluidity disappears, as appears to be observed experimentally
\cite{Rice2}.  Figure \ref{fig:n1n2}c reveals that when superfluidity is
present, it resides in the center of the trap, expelling the excess
fermions outside the core.
It should be noted that if one performs the same experiment as in
Fig.~\ref{fig:n1n2}a, but with much lower polarization (not shown), we
find that the crossing of the transition line will be barely observable.
as is necessary in order to be consistent with Fig.~\ref{fig:n1n2}b.  No
matter how the Sarma-PS transition line is crossed, at low $p$ (and thus
low $T$), the ratio $n_{\uparrow}/n_{\downarrow}$ remains 1. Therefore,
this class of experiments may not map out the transition curve for all
$p$.

Our theory can be generalized to address the whole of BEC-BCS crossover.
As one passes from unitarity towards the BEC regime we find the fraction
of the PS phase in the phase diagram decreases progressively,
disappearing at $1/k_Fa\approx 2.04$.  For stronger couplings, the
superfluid contribution to the phase diagram consists only of the Sarma
phase, at all $T$.  The same observation at $T=0$ was first reported in
Ref.~\cite{PS05a}. This can also be inferred from the homogeneous phase
diagram in Ref.~\cite{Chien06}, which shows that the Sarma state is
stable at low $T$ for any $p$ providing $1/k_Fa \gtrsim 2.3$.

\begin{figure}
\centerline{\includegraphics[clip,width=3.2in]{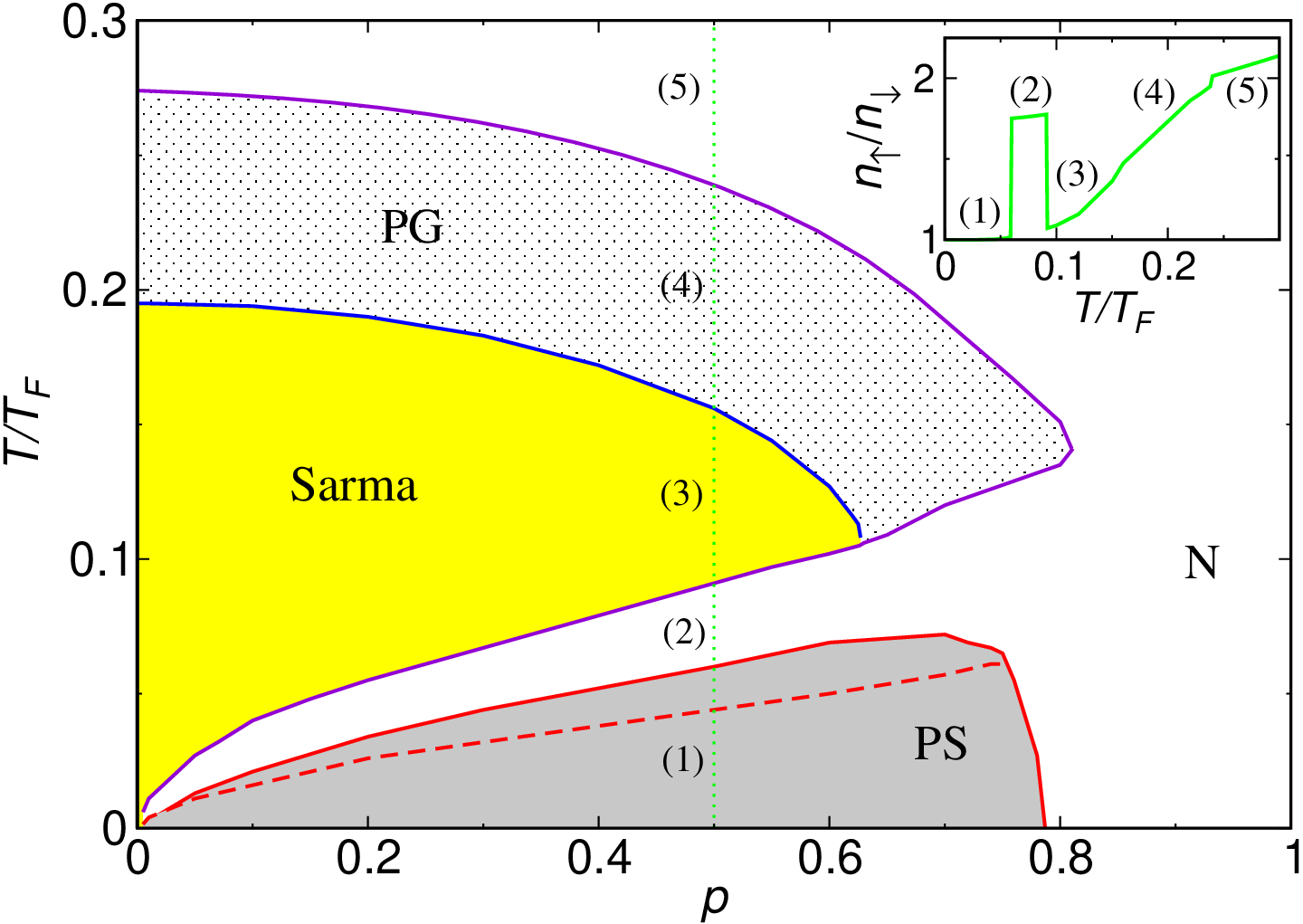}}
\caption{(Color online) Phase diagram of a polarized Fermi gas, as in
  Fig.~\ref{fig:TP_trap} except that $1/k_{F}a=-0.5$.  Here the Sarma
  and PS phases are separated by an intermediate normal regime.  The
  (green) dotted line indicates a sweep of $T$ at $p=0.5$, with five
  possible structures labeled: (1) PS, (2) N, (3) Sarma, (4) PG, (5) N.
  Shown in the inset is $n_{\uparrow}/n_{\downarrow}$ at the trap center
  as a function of $T/T_{F}$. }
\label{fig:TP_Ian05}
\end{figure}

In Fig.~\ref{fig:TP_Ian05} we present the phase diagram for a
population-imbalanced trapped gas on the BCS side of resonance, where
$1/k_{F}a=-0.5$.  It differs significantly from the unitary case and is,
in many ways even more rich.
Importantly, at $T=0$, the PS phase is no longer stable at very high
$p$.  We understand this by noting that in the BCS regime, the pairing
interaction is relatively weak and the gap $\Delta$ small. At
sufficiently high $p$, we have $h>\Delta(r=0)$ so that an unpolarized
BCS superfluid core can no longer be sustained.

Equally important is the fact that in this BCS case, the Sarma and PS
phases do not connect and an intermediate phase appears between the
Sarma and PS states.  As a consequence, the boundaries of the Sarma, PS,
and (pseudogapped or unpaired) normal phases do not meet except possibly
at $p=T=0$.  We presume that this intermediate phase is a normal Fermi
gas (N).
To understand its appearance, we note that as the BCS regime is
approached, (i) phase separation becomes problematic because finite
temperature (which enters via $T/\Delta$) has a stronger effect, allowing
the polarization to penetrate into the center of the core and thereby
making PS more difficult.  (ii) In addition, the intermediate
temperature Sarma phase becomes more fragile as the pairing is weakened.
As a result the Sarma and PS states retreat from each other as seen in
Fig.~\ref{fig:TP_Ian05}.  We cannot rule out LOFF-like states as an
alternative candidate in place of N. However, we have found \cite{LOFF1}
these phases in general have very low $T_c$ and should not persist at
these higher temperatures.  In the inset of Fig.~\ref{fig:TP_Ian05} we
plot $n_{\uparrow}/n_{\downarrow}$ at the trap center as a function of
$T/T_{F}$. The N state between the PS and Sarma phases would be
manifested by sudden jumps at the PS-N and N-Sarma boundaries. This
prediction can be used to test the existence of the intermediate N state
in the phase diagram.

We end with another prediction concerning how the ``mixed normal''
region of the trap, emphasized experimentally in Ref. \cite{ZSSK206},
evolves with $T$.  As noted in earlier work \cite{ChienRapid}, within
the Sarma phase, this mixed normal state consists of highly correlated
non-condensed pairs which necessarily become less significant as $T$
decreases.  However, with decreasing $T$, as seen from
Figs.~\ref{fig:TP_trap} and \ref{fig:TP_Ian05}, the Sarma phase gives
way to stable phase separation.  Only in the sliver above the (red)
dashed line will a phase separated gas contain a correlated mixed normal
region (appearing between the superfluid core and the domain wall).
Everywhere in the PS phase, there is a mixed normal region in the gas
outside the domain wall, with no pairing correlations, as appears at
strictly $T=0$ \cite{Mueller,HS06}.  Thus a change should occur from a
highly correlated to an uncorrelated mixed normal region at large radii
as $T$ is progressively decreased.  Further experiments should help to
address these predictions.

This work was supported by NSF PHY-0555325 and NSF-MRSEC Grant
No.~DMR-0213745; we thank Randy Hulet, Martin Zwierlein and Cheng Chin
for useful conversations.

\end{document}